\documentclass[preprint,sort&compress,1p]{elsarticle}
\usepackage{graphicx}
\usepackage{bm}
\usepackage{amsmath,amsfonts}
\usepackage[colorlinks=true]{hyperref}

\begin{document}
\title{Conserved momenta of a ferromagnetic soliton}

\author{Oleg Tchernyshyov}
\address{Department of Physics and Astronomy, 
	Johns Hopkins University,
	Baltimore, MD 21218, USA
}
\ead{olegt@jhu.edu}

\begin{abstract}
Linear and angular momenta of a soliton in a ferromagnet are commonly derived through the application of Noether's theorem. We show that these quantities exhibit unphysical behavior: they depend on the choice of a gauge potential in the spin Lagrangian and can be made arbitrary. To resolve this problem, we exploit a similarity between the dynamics of a ferromagnetic soliton and that of a charged particle in a magnetic field. For the latter, canonical momentum is also gauge-dependent and thus unphysical; the physical momentum is the generator of magnetic translations, a symmetry combining physical translations with gauge transformations. We use this analogy to unambiguously define conserved momenta for ferromagnetic solitons. General considerations are illustrated on simple models of a domain wall in a ferromagnetic chain and of a vortex in a thin film.
\end{abstract}

\maketitle

\section{Introduction}

The definition of linear and angular momenta of a ferromagnet is a subject with a long history, surprising results, and a lingering controversy \cite{Slonczewski1973, Thiele1976, Haldane1986, Volovik1987, Kosevich1990, Papanicolaou1991, Zhmudskii1999, Galkina2000, Sheka2006, Gaididei2010, Yan2013, Schutte2014}. It was realized in the 1970s that the linear momentum of a ferromagnetic soliton is determined not by its velocity but rather by its configuration \cite{Slonczewski1973, Thiele1976}. Although this sounds counterintuitive, one needs to realize that our physical intuition is based on experience with massive objects, for which an external force generates proportional acceleration. Spins in a ferromagnet behave differently: like fast-spinning gyroscopes, they precess at an angular velocity proportional to the external torque. An external force $\mathbf F$ acting during a short time interval $dt$  increments the velocity of a Newtonian particle, $\mathbf F dt = m \, d \mathbf v$. For a spin, an external torque $\bm \tau$ affects its orientation, $\bm \tau dt = d\mathbf S$. Thus the linear momentum of a ferromagnetic soliton is a function of its collective coordinates, rather than velocities.  

The earliest derivations of linear momentum in a ferromagnet followed the above qualitative reasoning and analyzed the configurational change of a soliton under a specified external perturbation. For example, a domain wall in a ferromagnetic chain pushed by an external magnetic field increments its azimuthal angle in proportion to the impulse of the force exerted by the field. The linear momentum of the domain wall is therefore proportional to its azimuthal angle \cite{Slonczewski1973}. 

On a deeper level, momenta are conserved quantities related to global symmetries. By Noether's theorem, invariance of the Lagrangian under translations and rotations gives rise to the conservation of linear and angular momenta. The precessional dynamics of the magnetization field $\mathbf m(\mathbf r,t)$ of unit length is represented in the Lagrangian not by a kinetic energy, but rather by a Berry-phase term $\mathcal L_B = \mathbf a \cdot \partial_t \mathbf m$. Here $\mathbf a(\mathbf m)$ is the gauge potential of a magnetic monopole \cite{Altland-Simons} whose field is  
\begin{equation}
\mathbf b(\mathbf m) = \nabla_{\mathbf m} \times \mathbf a(\mathbf m) 
	= - \mathcal J \mathbf m.
\label{eq:a-curl}
\end{equation} 
$\mathcal J$ is spin density. Eq.~(\ref{eq:a-curl}) does not fully specify $\mathbf a(\mathbf m)$: any gauge transformation 
\begin{equation}
\mathbf a'(\mathbf m) = \mathbf a(\mathbf m) + \nabla_\mathbf m \chi(\mathbf m) 
\label{eq:gauge-m}
\end{equation}
preserves $\mathbf b = \nabla_\mathbf m \times \mathbf a$. It is known that the linear momentum $\mathbf p$ derived from Noether's theorem is gauge-dependent \cite{Galkina2000, Yan2013}. This is a cause for concern: physical quantities should be gauge-invariant.

\begin{figure}
\includegraphics[width=0.9\columnwidth]{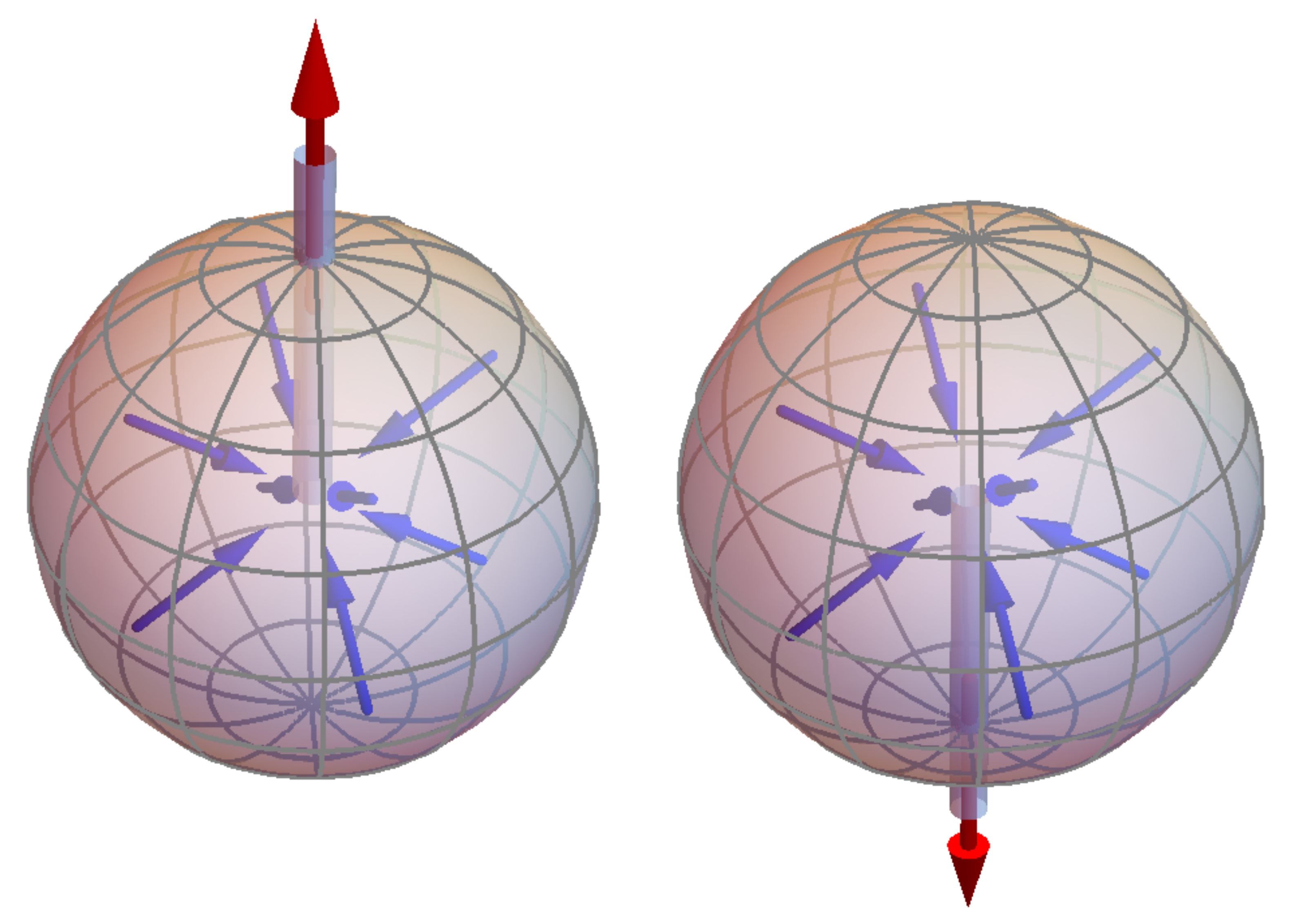}
\caption{Magnetic field $\mathbf b$ of a magnetic monopole with a net flux $-4\pi \mathcal J$ (blue arrows). A Dirac string at $\mathbf m_s$ (tube) carries an equal and opposite magnetic flux $+4\pi \mathcal J$ (red arrow). Standard choices are $\mathbf m_s = \pm \hat{\mathbf z}$.}
\label{fig:spheres1}
\end{figure}

A different but related problem arises for angular momentum $\mathbf J$. Although the magnetic field of a monopole (\ref{eq:a-curl}) is spherically symmetric, its gauge potential is not. The best we can do is to make it axially symmetric \cite{Haldane1986}:
\begin{equation}
\mathcal L_B = \mathbf a \cdot \partial_t \mathbf m 
	= \mathcal J 
		\frac{\mathbf m_s \times \mathbf m}{1 - \mathbf m_s \cdot \mathbf m} 
		 \cdot \partial_t \mathbf m.
\label{eq:gauge-generic}
\end{equation}
The singular direction $\mathbf m_s$ is the location of a Dirac string carrying away the magnetic flux $4\pi \mathcal J$, Fig.~\ref{fig:spheres1}. The axial symmetry of the Berry-phase term in the Lagrangian limits us to just one conserved component of angular momentum $\mathbf J \cdot \mathbf m_s$. The string is usually placed at $\mathbf m_s = \pm \hat{\mathbf z}$, which yields the familiar kinetic term \cite{Altland-Simons}
\begin{equation}
\mathcal L_B = \mathbf a \cdot \partial_t \mathbf m 
	= \mathcal J (\cos{\theta} \pm 1) \, \partial_t \phi.
\label{eq:gauge-poles}
\end{equation} 

We shall show below that linear and angular momenta defined through Noether's theorem for a Lagrangian with the kinetic term (\ref{eq:gauge-generic}) depend on the location of the string $\mathbf m_s$. Momenta defined in this way are essentially arbitrary, and thus unphysical, quantities. 

Our goal will be to define conserved momenta that are free from these problems. The resolution is tied to a powerful analogy between the dynamics of a magnetic soliton and of a charged particle in a magnetic field pointed out by Bar'yakhtar and Ivanov and by Papanicolaou \cite{Papanicolaou1991}. For a particle in a magnetic field, the gauge potential $\mathbf A$ breaks translational invariance even if the field $\mathbf B = \nabla \times \mathbf A$ itself is uniform. Canonical momentum $\mathbf p = \partial L/\partial \dot{\mathbf r} = m \dot{\mathbf r} + e \mathbf A$ is neither gauge invariant, nor conserved. The proper symmetries of the Lagrangian are combinations of translations and gauge transformations known as magnetic translations \cite{Zak1964}. Their generator 
\begin{equation}
\mathbf P = \mathbf p - e \mathbf A - e \mathbf r \times \mathbf B = m \dot{\mathbf r} - e \mathbf r \times \mathbf B
\label{eq-P-intro}
\end{equation}
is both gauge-invariant and conserved. A related quantity, the guiding center of the cyclotron orbit $\mathbf r_0$, is defined by the equation $\dot{\mathbf r} = (\mathbf r - \mathbf r_0) \times e \mathbf B/m$ \cite{Landau-Lifshitz-III}. 

Conserved momentum of a particle $\mathbf P = m \dot{\mathbf r} - e \mathbf r \times \mathbf B$ has a transparent physical meaning. Newton's second law contains three terms, 
\begin{equation}
m \ddot{\mathbf r} = e \dot{\mathbf r} \times \mathbf B + \mathbf F, 
\label{eq:Newton-2nd-law}
\end{equation}
where the Lorentz force $e \dot{\mathbf r} \times \mathbf B$ has been set aside. Treating it as a kinematic effect, just like the inertia term $m \ddot{\mathbf r}$, gives the right perspective. The impulse of the external force $\mathbf F$ then defines the change in momentum. 
\begin{equation}
\Delta \mathbf P \stackrel{\mathrm{def}}{=} \int \mathbf F \, dt = \Delta (m \dot{\mathbf r} - e \mathbf r \times \mathbf B).
\label{eq:magnetic-momentum}
\end{equation}
Hence Eq.~(\ref{eq-P-intro}).

Below we will first show that the momenta of a ferromagnetic soliton derived through the na\"{\i}ve application of Noether's theorem are  canonical momenta conjugate to the soliton's zero modes. They are neither gauge invariant, nor conserved, generally speaking. We will then derive the physical conserved momenta that are analogs of the generator of magnetic translations. The method is applicable to any magnetic soliton and is not restricted to the symmetries of translation and rotation.

The paper is organized as follows. Sec.~\ref{sec:mechanics} introduces the general formalism of Lagrangian mechanics for solitons in a ferromagnet and defines gauge-dependent canonical momenta and their gauge-invariant conserved counterparts. The formalism is illustrated on the examples of a domain wall in one dimension (Sec.~\ref{sec:domain-wall}) and of a vortex in two dimensions (Sec.~\ref{sec:vortex}). Sec.~\ref{sec:discussion} summarizes the main results. 

\section{Mechanics of a soliton}
\label{sec:mechanics}

\subsection{Lagrangian and the gauge field}
\label{sec:mechanics-Lagrangian}

A soliton can be described in terms of some collective coordinates $\{q_1, q_2, \ldots\}$ such as its position, size, orientation, shape, and so on \cite{Tretiakov2008}. Time evolution of the magnetization field $\mathbf m(\mathbf r,t)$ can then be expressed as the evolution of these coordinates: $\partial_t \mathbf m = \dot{q}_i \, \partial \mathbf m/\partial q_i$. Here $\dot{q}_i = dq_i/dt$; summation is implied over doubly repeated indices.  The Lagrangian $L = \int \mathbf a \cdot \partial_t \mathbf m \, dV - U[\mathbf m]$, expressed in terms of  collective coordinates, becomes \cite{Clarke2008}
\begin{equation}
L = A_i(\mathbf q) \dot{q}_i - U(\mathbf q), 
\quad 
A_i = \int dV \,  \mathbf a(\mathbf m) \cdot \partial \mathbf m/\partial q_i.
\label{eq:L-soliton}
\end{equation}
The Lagrangian is similar to that of a massless particle moving in an external magnetic field, whose strength $F_{ij}$ is given by the curl of the gauge potential $A_i$:
\begin{equation}
F_{ij} = \partial_i A_j - \partial_j A_i. 
\label{eq:F-def}
\end{equation}
$U$ is potential energy and $\partial_i \equiv \partial/\partial q_i$. 

To prevent confusion, let us note that the Lagrangian with no mass terms (\ref{eq:L-soliton}) applies when we keep \emph{all} of the system's collective coordinates, not just its soft modes. If we integrate out some hard modes, soft modes may acquire inertia. This is the origin of D\"{o}ring's mass \cite{Doring1948, Becker1952} observed in domain walls \cite{Saitoh2004} and skyrmions \cite{Buttner2015}.

To make contact with the more familiar case of a massive particle, we will add an inertia term, 
\begin{equation}
L = m \dot{q}_i \dot{q}_i/2 +  A_i \dot{q}_i - U, 
\label{eq:L-particle}
\end{equation}
and will set $m = 0$ when we deal with magnetic solitons.

\subsection{Canonical momenta and conserved momenta}
\label{sec:mechanics-momenta}

Continuous symmetries are associated with cyclic coordinates, which do not affect the potential energy and the magnetic field: 
\begin{equation}
\partial_a U = 0, 
\quad
\partial_a F_{ij} = 0. 
\label{eq:cyclic-coordinates}
\end{equation}
To distinguish these cyclic coordinates from the rest, we reserve for them indices from the beginning of the alphabet, $a, b, \ldots$  Canonical momenta conjugate to cyclic coordinates are
\begin{equation}
p_a = \partial L/\partial \dot{q}_a = m \dot{q}_a + A_a. 
\label{eq:p-canonical}
\end{equation}
They are neither gauge invariant, nor conserved: $\dot{p}_a = \partial_a A_i \, \dot{q}_i \neq 0$ for a generic gauge choice. 

To construct conserved momenta, we examine the transformation of the Lagrangian under an infinitesimal translation: 
\begin{subequations}
\begin{equation}
\delta q_a = \epsilon_a, 
\quad
\delta A_i = \epsilon_a \partial_a A_i, 
\quad
\delta L = \dot{q}_i \delta A_i. 
\label{eq:translation}
\end{equation}
The change in the Lagrangian stems from the lack of translational invariance of the gauge potential, $\mathbf A(\mathbf q + \bm \epsilon) \neq A(\mathbf q)$. However, the translated potential $\mathbf A(\mathbf q + \bm \epsilon)$ describes the same magnetic field, $\mathbf F(\mathbf q + \bm \epsilon) = \mathbf F(\mathbf q)$, by virtue of translational symmetry (\ref{eq:cyclic-coordinates}). We may then follow up the translation (\ref{eq:translation}) with a gauge transformation
\begin{equation}
\delta A_i = \partial_i \chi(\mathbf q),
\quad
\delta L = \dot{q}_i \, \partial_i \chi = d\chi/dt,
\label{eq:gauge-transform}
\end{equation}
\end{subequations}
to return the gauge field, and with it the Lagrangian, to its original form. (Adding a full time derivative to the Lagrangian does not affect the equations of motion.) Conserved momentum is the generator of the combined symmetry of translation and gauge transformation,
\begin{equation}
P_a = \partial L/\partial \dot{q}_a + \partial \chi/\partial \epsilon_a.
\label{eq:P-def}
\end{equation}

To find the right gauge transformation (\ref{eq:gauge-transform}), we solve the equation $\partial_i \chi + \epsilon_a \partial_a A_i = 0$ for $\chi$. To that end, we rewrite $\partial_a A_i = F_{ai} + \partial_i A_a$ and obtain a tentative answer 
\begin{equation}
\chi(\mathbf q) = - \epsilon_a A_a(\mathbf q) - \epsilon_a \int F_{ai}(\mathbf q) \, dq_i. 
\label{eq:chi-answer}
\end{equation}
It remains to verify that the second term in Eq.~(\ref{eq:chi-answer}) depends on the initial and finite positions but not of the integration path between them. That would be the case if the integral over any closed loop vanishes, $\oint F_{ai} \, dq_i = 0$. In the differential form, $\partial_j F_{ai} - \partial_i F_{aj} = 0$ (zero curl). That this condition is fulfilled can be seen by noting that the field is uniform in the $q_a$ direction, $\partial_a F_{ij} = 0$, and that it satisfies the Jacobi identity 
\begin{equation}
\partial_i F_{jk} + \partial_j F_{ki} + \partial_k F_{ij} = 0
\label{eq:Jacobi}
\end{equation}
as long as it is derived from a gauge potential, $F_{ij} = \partial_i A_j - \partial_j A_i$. Upon combining Eqs.~(\ref{eq:P-def}) and (\ref{eq:chi-answer}), we obtain the conserved momentum
\begin{equation}
P_a(\mathbf q, \dot{\mathbf q}) = p_a - A_a - \int F_{ai} \, dq_i
	= m \dot{q}_a - \int F_{ai} \, dq_i.
\label{eq:P-particle}
\end{equation}
This quantity is both gauge invariant and conserved. Eq.~(\ref{eq-P-intro}) is a particular case of a particle in 3 dimensions in a uniform magnetic field, $F_{ab} = \epsilon_{abc} B_c$.

An alternative way to derive the conserved momentum is to start with Newton's second law, 
\begin{equation}
m \ddot{q}_a = F^{(L)}_a + F^{(c)}_a,
\label{eq:2nd-law}
\end{equation}
where the right-hand side includes the Lorentz force $F^{(L)}_a = F_{ai} \dot{q}_i$ and the conservative force $F^{(c)}_a = - \partial_a U$. For a cyclic coordinate, the conservative force vanishes. Let us break the symmetry and apply an external force $F^{(c)}_a \neq 0$. Its impulse induces a change in the momentum $P_a$. With the aid of Eq.~(\ref{eq:2nd-law}), we obtain
\[
\Delta P_a = \int F^{(c)}_a dt = m \Delta \dot{q}_a - \int F_{ai} dq_i, 
\]
in agreement with Eq.~(\ref{eq:P-particle}). Thiele \cite{Thiele1976} followed similar logic to define linear momentum of a domain wall.

For a soliton in a ferromagnet, the mass $m = 0$, so conserved momenta are functions of  coordinates:
\begin{equation}
P_a(\mathbf q_2) - P_a(\mathbf q_1) 
	= - \int_{\mathbf q_1}^{\mathbf q_2} F_{ai}(\mathbf q) \, dq_i.
\label{eq:P-soliton}
\end{equation}
The role of the magnetic field $F_{ij}$ is played by the gyrotropic tensor for collective coordinates \cite{Clarke2008}
\begin{eqnarray}
F_{ij} 
 	&=& \mathcal J \int \mathbf m \cdot 
		\left( 
			\frac{\partial \mathbf m}{\partial q_i}
			\times 
			\frac{\partial \mathbf m}{\partial q_j}
		\right) \, 
		dV
\nonumber\\
	&=& \mathcal J \int 
		\left(
			\frac{\partial \phi}{\partial q_i} \frac{\partial \cos{\theta}}{\partial q_j}
			-  \frac{\partial \phi}{\partial q_j}
				\frac{\partial \cos{\theta}}{\partial q_i}
		\right) \, dV.
\label{eq:F-gyrotropic}
\end{eqnarray}
This expression can also be obtained directly from the definitions of the gauge potential (\ref{eq:L-soliton}) and magnetic field (\ref{eq:F-def}), see \ref{appendix:F-from-A}. 

It is useful to translate the definition of momentum (\ref{eq:P-soliton}) from the language of collective coordinates back to the field-theoretic description. To compute the difference of momenta $P_a$ between two arbitrary states $\mathbf m_1(\mathbf r)$ and $\mathbf m_2(\mathbf r)$ of a magnetic soliton, we imagine evolving it in time from the initial state $\mathbf m(\mathbf r,t_1) = \mathbf m_1(\mathbf r)$ to the final state $\mathbf m(\mathbf r,t_2) = \mathbf m_2(\mathbf r)$. We express $F_{ai} \, dq_i = F_{ai} \, \dot{q}_i dt$ with the aid of Eq.~(\ref{eq:F-gyrotropic}) to obtain
\begin{eqnarray}
&& P_a[\mathbf m_2(\mathbf r)] - P_a[\mathbf m_1(\mathbf r)] 
\nonumber\\
&& = 
	- \mathcal J 
		\int_{t_1}^{t_2} dt \int dV \, 
			\mathbf m \cdot 
			\left( 
				\frac{\partial \mathbf m}{\partial q_a}
				\times
				\frac{\partial \mathbf m}{\partial t} 
			\right).
\label{eq:P-field}
\end{eqnarray}
The momentum difference depends only on the initial and final configurations but not on the path between them, nor on how fast the evolution happens. 

Canonical momentum is 
\begin{equation}
p_a = \partial L/\partial \dot{q}_a = A_a 
	= \int dV \,  \mathbf a(\mathbf m) \cdot \frac{\partial \mathbf m}{\partial q_i}.
\label{p-canonical-soliton}
\end{equation}

\subsection{Some useful relations}
\label{sec:mechanics-stuff}

The standard Poisson bracket is defined as
\begin{equation}
\{f,g\} = \sum_{i} 
	\left( 
		\frac{\partial f}{\partial p_i} \frac{\partial g}{\partial q_i}
		- \frac{\partial f}{\partial q_i} \frac{\partial g}{\partial p_i}
	\right).
\label{eq:Poisson-particle-massive}
\end{equation}
Poisson brackets for coordinates and conserved momenta of a massive particle are thus
\begin{equation}
\{q_i, q_j\} = 0,
\quad
\{P_a, q_i\} = \delta_{ai},
\quad
\{P_a, P_b\} = - F_{ab}.
\label{eq:Poisson-particle}
\end{equation}

For magnetic solitons, the mass term vanishes so that canonical momenta (\ref{eq:p-canonical}) lose their dependence on velocities and become functions of coordinates only, $p_a = A_a(\mathbf q)$. Because coordinates and canonical momenta are no longer independent, the standard definition of the Poisson bracket (\ref{eq:Poisson-particle-massive}) breaks down and needs to be modified. 

In a ferromagnet, the $z$ component of a spin $\hbar S \cos{\theta}$ is the canonical momentum conjugate to its azimuthal angle $\phi$. Hence the Poisson bracket of a discrete set of spins $\{\mathbf S_{\mathbf r}\}$ \cite{Mermin1964},
\begin{equation}
\{f,g\} = \frac{1}{\hbar S} \sum_{\mathbf r} 
	\left( 
		\frac{\partial f}{\partial \cos{\theta_{\mathbf r}}} 
			\frac{\partial g}{\partial \phi_{\mathbf r}}
		- \frac{\partial g}{\partial \cos{\theta_{\mathbf r}}} 
			\frac{\partial f}{\partial \phi_{\mathbf r}} 
	\right),
\end{equation}
and the continuum version, 
\begin{equation}
\{f,g\} = \frac{1}{\mathcal J} \int 
	\left( 
		\frac{\delta f}{\delta \cos{\theta}} \frac{\delta g}{\delta \phi}
		- \frac{\delta g}{\delta \cos{\theta}} \frac{\delta f}{\delta \phi} 
	\right) \, dV.
\end{equation}
The Poisson brackets for collective coordinates and conserved momenta of a ferromagnet are (\ref{appendix:Poisson})
\begin{equation}
\{q_i, q_j\} = (F^{-1})_{ij},
\ 
\{P_a, q_i\} = -\delta_{ai},
\ 
\{P_a, P_b\} = - F_{ab}.
\label{eq:Poisson-soliton}
\end{equation}

Eq.~(\ref{eq:Poisson-soliton}) indicates that at the quantum level conserved momenta of a soliton do not commute if the corresponding gyrotropic coefficient is nonzero. Watanabe and Murayama \cite{Watanabe2014} pointed this out for a ferromagnetic skyrmion. We see here that non-commutativity of conserved momenta in a ferromagnet is a generic feature. Furthermore, coordinates are non-commutative, too!

The following gauge-invariant relations will be helpful: 
\begin{subequations}
\begin{eqnarray}
\frac{\partial p_b}{\partial q_a} - \frac{\partial p_a}{\partial q_b} &=& F_{ab},\\
\frac{\partial P_b}{\partial q_a} - \frac{\partial P_a}{\partial q_b} &=& 2F_{ab}.
\end{eqnarray}
They follow directly from the definitions of conserved (\ref{eq:P-soliton}) and canonical (\ref{p-canonical-soliton}) momenta of a soliton. Eqs.~(\ref{eq:p-q-relations}) indicate that it is impossible to find a gauge in which two canonical momenta are equal to their conserved counterparts,  $p_a = P_a$ and $p_b = P_b$, unless the gyrotropic coefficient $F_{ab}$ vanishes. 
\label{eq:p-q-relations}
\end{subequations}

\section{Domain wall in one dimension} 
\label{sec:domain-wall}

\subsection{Lagrangian and soliton solutions}
\label{sec:domain-wall-solutions}

\begin{figure}
\includegraphics[width=0.95\columnwidth]{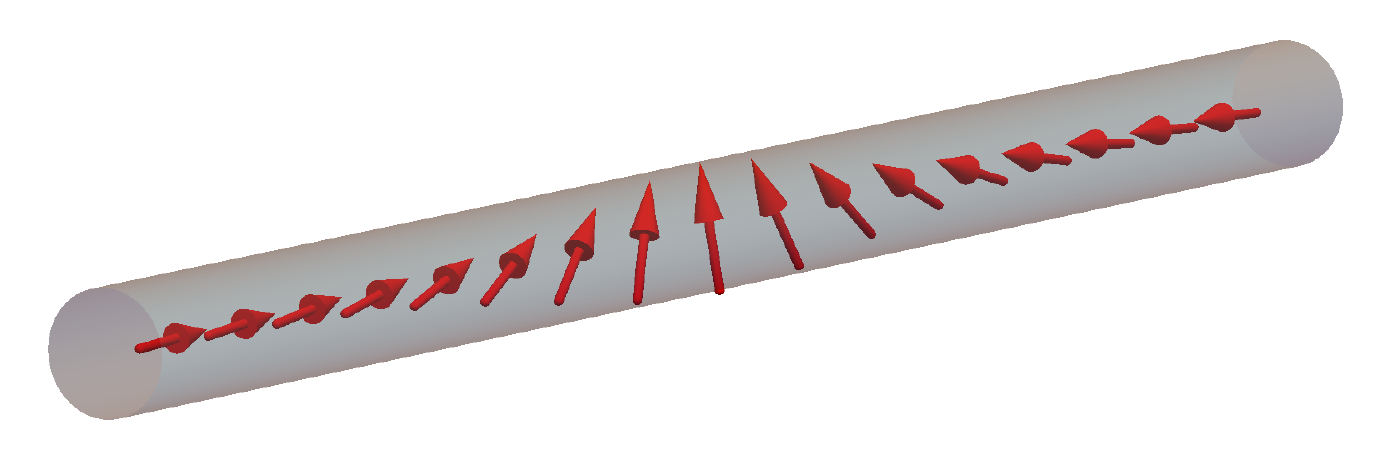}
\caption{A domain wall in a ferromagnetic wire.}
\label{fig:domain-wall}
\end{figure}

A ferromagnetic chain with easy-axis anisotropy has a Lagrangian density
\begin{equation}
\mathcal L = \mathbf a(\mathbf m) \cdot \partial_t \mathbf m 
	- \frac{A(\partial_z \mathbf m)^2}{2} 
	- \frac{K(\mathbf m \times \hat{\mathbf z})^2}{2}.
\end{equation}
The system has two uniform ground states $\mathbf m(z) = \pm \hat{\mathbf z}$. We shall consider here topological solitons satisfying the following boundary conditions: 
\begin{equation}
\mathbf m(\pm \infty) = \pm \hat{\mathbf z}.
\label{eq:domain-wall-bc}
\end{equation}

In equilibrium, a domain wall has the width $\lambda = \sqrt{A/K}$ and configuration
\begin{equation}
m_x + i m_y = e^{i\Phi} \, \mathrm{sech}\frac{z-Z}{\lambda},
\quad
m_z = \tanh{\frac{z-Z}{\lambda}}. 
\label{eq:domain-wall-equilibrium}
\end{equation}
The two free parameters---the location of the wall $Z$ and the azimuthal angle $\Phi$---are zero modes associated with the symmetries of translation and spin rotation about the $z$ axis. 

\begin{figure}
\includegraphics[width=0.9\columnwidth]{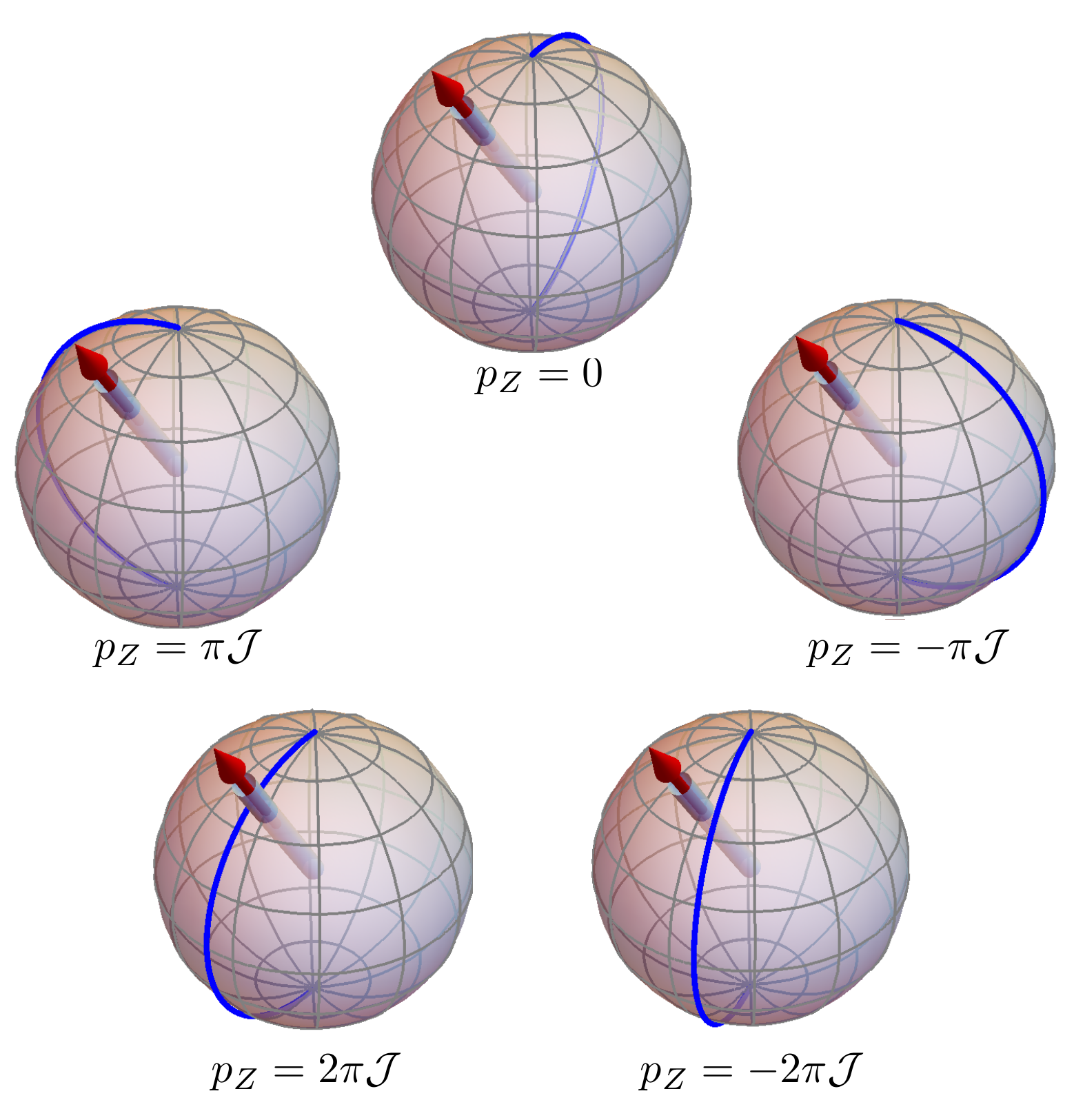}
\includegraphics[width=0.9\columnwidth]{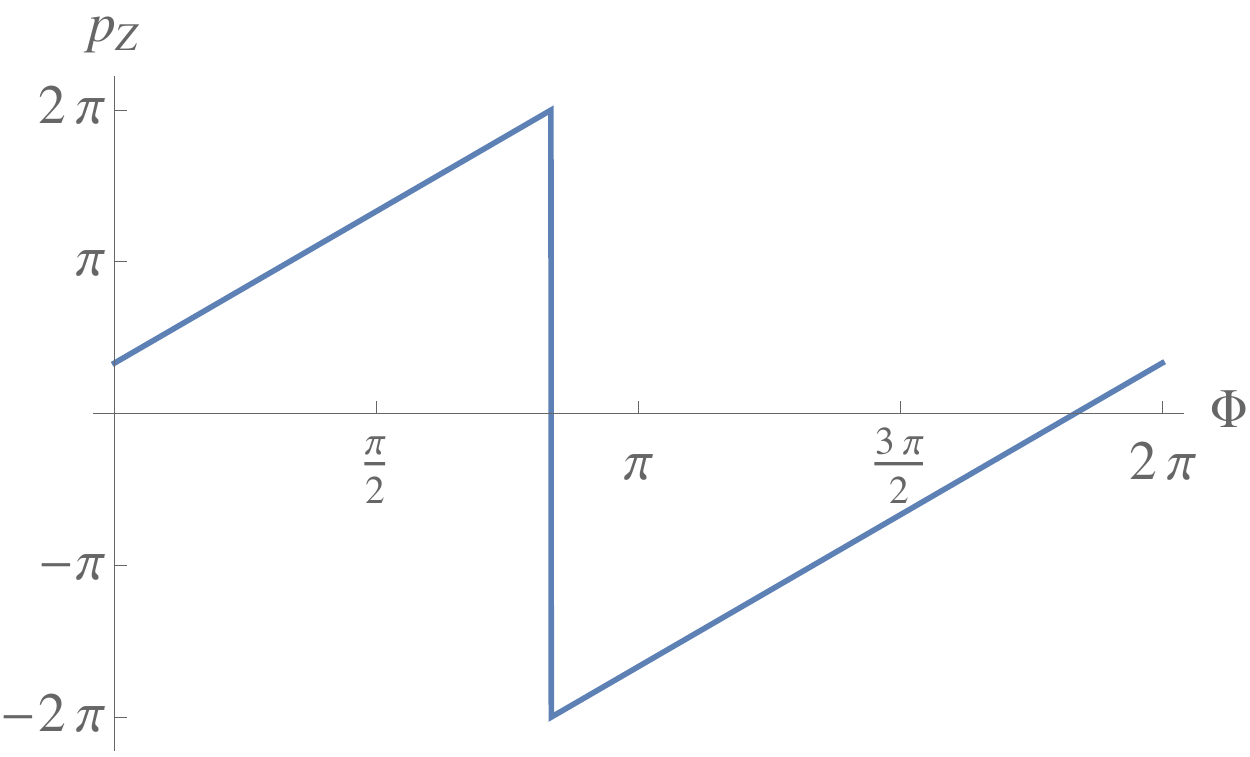}
\caption{A one-dimensional domain wall in equilibrium (\ref{eq:domain-wall-equilibrium}) is depicted on the unit sphere of magnetization $\mathbf m$ as an arc connecting the south and north poles along the $\Phi$ meridian. With the Dirac string (tube) away from the poles, $\mathbf m_s \neq \pm \hat{\mathbf z}$, canonical momentum $p_Z$ (\ref{eq:pZ-wall-generic}) increases linearly with $\Phi$ as the arc moves east, from $p_Z = -2\pi \mathcal J$ just east of the string to $+2\pi \mathcal J$ just west of the string.}
\label{fig:spheres2}
\end{figure}

\subsection{Canonical momenta}
\label{sec:domain-wall-p}

Canonical linear momentum is
\begin{equation}
p_Z = - \int_{-\infty}^{+\infty} a(\mathbf m) \cdot \partial_z \mathbf m \, dz
	= - \int_{-\hat{\mathbf z}}^{+\hat{\mathbf z}} a(\mathbf m) \cdot d \mathbf m,
\label{eq:p-canonical-wall}
\end{equation}
with the integration path between the poles taken along the meridian $\Phi = \mathrm{const}$. In deriving this result from Eq.~(\ref{p-canonical-soliton}) we relied on the fact that $\partial/\partial Z = - \partial/\partial z$ for a rigidly translated domain wall $\mathbf m(z-Z)$. The expression for canonical momentum (\ref{eq:p-canonical-wall}) coincides with that derived through the application of Noether's theorem \cite{Galkina2000, Yan2013}. It changes under a gauge transformation (\ref{eq:gauge-m}):
\begin{equation}
p_Z' = p_Z - \int_{-\hat{\mathbf z}}^{+\hat{\mathbf z}} 
	\nabla_{\mathbf m}\chi(\mathbf m) \cdot d \mathbf m
	= p_Z - \chi(+\hat{\mathbf z}) + \chi(-\hat{\mathbf z}).
\end{equation}
We see that canonical momentum $p_Z$ of a domain wall $\mathbf m(\mathbf r)$ is not well defined.  

We may attempt to remove the unphysical gauge dependence by evaluating the \emph{difference} of momenta between two states of a domain wall $\mathbf m_1(z)$ and $\mathbf m_2(z)$ \cite{Galkina2000, Yan2013}. The gauge dependence cancels out, provided that the two configurations are in the same topological sector (\ref{eq:domain-wall-bc}).

However, we are not out of the woods yet. Although momentum difference $p_Z[\mathbf m_2(z)] - p_Z[\mathbf m_2(z)]$ does not change after a gauge transformation, this quantity is still sensitive to the location of the Dirac string $\mathbf m_s$ in Eq.~(\ref{eq:gauge-generic}) as we shall see next. As the string is an artifact of the gauge description, the dependence on its location makes canonical linear momentum unphysical even after the subtraction. 

\begin{subequations}
For an axially symmetric gauge (\ref{eq:gauge-generic}) with the Dirac string away from the poles, we obtain 
\begin{eqnarray}
p_Z &=& - 4 \mathcal J \arctan{\cot{\frac{\Phi - \phi_s}{2}}}
\nonumber\\
	&=& 2 \mathcal J [\Phi - \phi_s - (2n+1)\pi], 
\quad  
\mathbf m_s \neq \pm \hat{\mathbf z},
\label{eq:pZ-wall-generic}
\end{eqnarray}
for $\Phi$ between $\phi_s + 2 n \pi$ and $\phi_s + 2(n+1)\pi$. 
Momentum increases linearly with $\Phi$ from $p_Z = - 2\pi \mathcal J$ with the domain wall just east of the Dirac string to $p_Z = + 2\pi \mathcal J$ just west of the string, Fig.~\ref{fig:spheres2}. Crossing the string results in a jump $\Delta p_Z = -4\pi \mathcal J$. Thus the presence of the string ensures the single-valuedness of canonical momentum. 

The $4\pi \mathcal J$ discontinuity of canonical momentum is a bug in the classical field theory of a ferromagnet. It goes away when we recall that the continuum theory is derived from a system with discrete quantum spins \cite{Haldane1986}. Translations become a discrete symmetry. In a chain with a lattice spacing $a$ and spins of length $S$, the spin density is $\mathcal J = \hbar S/a$. The generator of translations is the translation by one lattice spacing, $T(a) = \exp{(- i p_Z a/\hbar)}$. Adding $4\pi \mathcal J$ to $p_Z$ multiplies the translation operator by
\begin{equation}
\exp{\left( - 4 \pi i \mathcal J a/\hbar\right)} 
	= \exp{\left( - 4 \pi i S\right)} = 1
\end{equation}
if $S$ is integer or half-integer. We see that the discontinuity does not affect the physical operator $T(a)$. 

Although the appeal to the discrete and quantum nature of the ferromagnet saves the day, the solution comes at a high price of abandoning the classical field theory. There are other problems, too.

When the Dirac string is located at one of the poles, $\mathbf m_s = \pm \hat{\mathbf z}$, Eq.~(\ref{eq:pZ-wall-generic}) does not apply as the azimuthal coordinate of the string $\phi_s$ is undefined. Evaluation of Eq.~(\ref{eq:p-canonical-wall}) in the gauge (\ref{eq:gauge-poles}) yields 
\begin{equation}
p_Z = 0, 
\quad
\quad  
\mathbf m_s = \pm \hat{\mathbf z},
\label{eq:pZ-wall-poles}
\end{equation}
because $d\phi = (\partial \phi/\partial z) dz = 0$ in equilibrium, see Eq.~(\ref{eq:domain-wall-equilibrium}). 
\label{eq:p-Z}
\end{subequations}
Linear momentum vanishes in this gauge.

Canonical angular momentum $p_\Phi$ is equally problematic. To evaluate it, we introduce a local reference frame with mutually orthogonal unit vectors pointing up, south, and east:
\begin{equation}
\mathbf e_1 = \mathbf m,
\quad
\mathbf e_2 = \frac{\partial \mathbf m}{\partial \theta},
\quad
\mathbf e_3 = \frac{1}{\sin{\theta}} \frac{\partial \mathbf m}{\partial \phi}.
\end{equation}
With the aid of Eq.~(\ref{p-canonical-soliton}) we obtain 
\[
p_\Phi = - \mathcal J \int_{-\infty}^{+\infty} 
	\frac{\mathbf m_s \cdot \mathbf e_2 \, \sin{\theta}}{1 - \mathbf m_s \cdot \mathbf e_1} 
	\, dz.
\]
For a domain wall in equilibrium (\ref{eq:domain-wall-equilibrium}), $- \mathbf e_2 \sin{\theta} = \lambda \, \partial \mathbf e_1/ \partial z$, which allows us to evaluate the integral,
\begin{equation}
p_\Phi = - \mathcal J \lambda 
	\left. \ln{[1 - \mathbf m_s \cdot \mathbf m(z)]} \right|_{-\infty}^{+\infty}.
\label{eq:p-Phi-canonical}
\end{equation}

For a generic location of the Dirac string away from the poles, canonical angular momentum is a gauge-dependent constant:
\begin{subequations}
\begin{equation}
p_\Phi = 2\mathcal J \lambda \ln{\cot{\frac{\theta_s}{2}}},
\quad
\mathbf m_s \neq \pm \hat{\mathbf z}.
\end{equation}
With the string at one of the poles, the result diverges and we need to work with a chain of large but finite length $L \gg \lambda$. Working in a standard gauge (\ref{eq:gauge-poles}) yields 
\[
p_\Phi = \mathcal J \int_{-L/2}^{+L/2} (\cos{\theta} \pm 1) \, dz.
\]
The same expression can be obtained via Noether's theorem \cite{Zhmudskii1999, Gaididei2010}. For a long chain, $L \gg |Z|$, $L \gg \lambda$, 
\begin{equation}
p_\Phi = \pm \mathcal J L  - 2 \mathcal J Z,
\quad
\mathbf m_s = \pm \hat{\mathbf z}.
\end{equation}
\label{eq:p-Phi}
\end{subequations}

Like its linear counterpart, canonical angular momentum shows strong gauge dependence. In the standard gauges (\ref{eq:gauge-poles}), $p_\Phi$ depends on position $Z$. For other positions of the Dirac string, it does not.

The dependence of canonical momenta (\ref{eq:p-Z}) and (\ref{eq:p-Phi}) on the location of the Dirac string clearly makes them unphysical. It is not obvious a priori which of the answers, if any, is correct. Our best bet is to avoid them entirely and to use conserved momenta that are free from these artifacts.

\subsection{Conserved momenta}
\label{sec:domain-wall-P}

Gauge-invariant conserved momenta $P_Z$ and $P_\Phi$ for a domain wall in equilibrium can be computed with little effort. Because no other (hard) modes are excited, we only need the gyrotropic coefficients (\ref{eq:F-gyrotropic}) involving both soft modes, $F_{\Phi Z} = - F_{Z \Phi} = 2 \mathcal J$ \cite{Clarke2008}. Using the domain wall with $Z = \Phi = 0$ as a reference point, we obtain with the aid of Eq.~(\ref{eq:P-soliton}) 
\begin{eqnarray}
P_Z(Z,\Phi) &=& P_Z(0,0) + 2 \mathcal J \Phi, 
\label{eq:P-Z-eq}
\\
P_\Phi(Z,\Phi) &=& P_\Phi(0,0) - 2 \mathcal J Z. 
\label{eq:P-Phi-eq}
\end{eqnarray}
The proportionality of angular momentum $P_\Phi$ to coordinate $Z$ is easy to understand.  In a ferromagnetic chain, angular momentum comes from spin alone. Shifting the domain wall from 0 to $Z$ elongates the $\mathbf m = - \hat{\mathbf z}$ domain and shortens the $\mathbf m = + \hat{\mathbf z}$ domain by $Z$, thereby reducing the $z$ component of spin by $2 \mathcal J Z$. This result is not sensitive to the detailed structure of the domain wall in equilibrium (\ref{eq:domain-wall-equilibrium}) and remains valid as long as the soliton interpolates between the two ground states, $\mathbf m(\pm \infty) = \pm \hat{\mathbf z}$, and the azimuthal angle is spatially uniform, $\phi(z) = \Phi$. 

Unlike its canonical counterpart, conserved momentum $P_Z$ (\ref{eq:P-Z-eq}) is not single-valued: advancing the azimuthal angle $\Phi$ by $2\pi$ increments $P_Z$ by $4\pi \mathcal J$ even though the domain wall returns to the original state. The multi-valuedness is connected to the absence of the Dirac string in the gauge-invariant treatment and to the ensuing violation of the Jacobi identity (\ref{eq:Jacobi}). 

Comparing the canonical momenta obtained in various gauges (\ref{eq:p-Z}) and (\ref{eq:p-Phi}) with the physical answers (\ref{eq:P-Z-eq}) and (\ref{eq:P-Phi-eq}), we observe that canonical momenta $p_a$ sometimes reproduce the correct answers $P_a$ and sometimes they do not. Furthermore, no gauge choice yields $p_Z = P_Z$ \emph{and} $p_\Phi = P_\Phi$. The canonical approach gets right at most one or the other, but not both. That is not a coincidence: Eqs.~(\ref{eq:p-q-relations}) show that, as long as $F_{ab} \neq 0$, there is no gauge in which two canonical momenta $p_a$ and $p_b$ can match their physical counterparts $P_a$ and $P_b$. 

One may wonder how Yan \emph{et al.} \cite{Yan2013} managed to get both linear and angular momenta right in the same gauge. The answer is their linear momentum should have been 0 as in Eq.~(\ref{eq:pZ-wall-poles}). As explained by Thiele \cite{Thiele1976}, the formula $p_Z = \mathcal J \int (\cos{\theta} \pm 1) \, d\phi$ in a standard gauge (\ref{eq:gauge-poles}) is missing crucial boundary terms. Yan \emph{et al.} \cite{Yan2013} smuggled them in. 

To compute the linear momentum of a domain wall with arbitrary deformations, we use Eq.~(\ref{eq:P-field}) and replace $\partial / \partial Z$ with $-\partial / \partial z$ (as $Z$ represents a rigid displacement) to obtain
\begin{eqnarray}
 P_Z[\mathbf m_2(z)] - P_Z[\mathbf m_1(z)] 
 = 
	\mathcal J 
		\iint  
			\mathbf m \cdot 
			\left( 
				\frac{\partial \mathbf m}{\partial z}
				\times
				\frac{\partial \mathbf m}{\partial t} 
			\right) dt \, dz
= \mathcal J \mathcal A.
\label{eq:P-1-d}
\end{eqnarray}
The double integral represents the area $\mathcal A$ swept by the curve $\mathbf m(z,t)$ on the unit sphere as it evolves from $\mathbf m_1(z)$ to $\mathbf m_2(z)$. This result was correctly anticipated by Galkina and Ivanov \cite{Galkina2000} and by Yan \emph{et al.} \cite{Yan2013}. If both the initial and final states are equilibrium configurations (\ref{eq:domain-wall-equilibrium}) then the momentum difference is $2 \mathcal J (\Phi_2 - \Phi_1)$, in agreement with Eq.~(\ref{eq:P-Z-eq}). Conserved linear momentum for a domain wall (\ref{eq:P-1-d}) was first obtained by Thiele \cite{Thiele1976}, who derived it by integrating the impulse of the gyrotropic force. 

\section{Vortex in a thin film} 
\label{sec:vortex}

\subsection{Soliton solutions}
\label{sec:vortex-solitons}

\begin{figure}
\includegraphics[width=0.98\columnwidth]{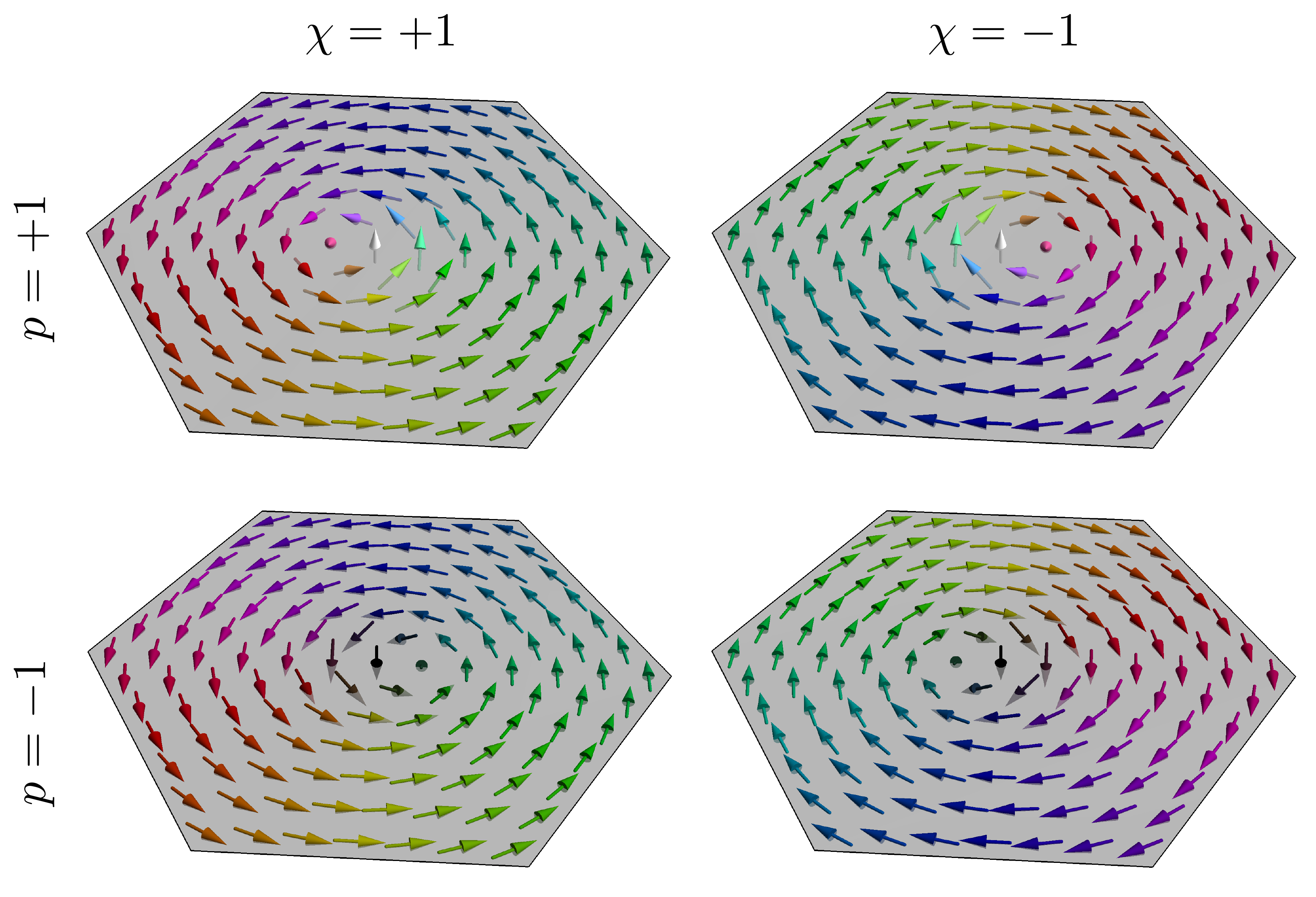}
\caption{Four possible equilibrium configurations of a vortex in a thin film. Arrows represent the in-plane components $m_x$ and $m_y$, their colors the azimuthal magnetization angle $\phi$, and their brightness the out-of-plane component $m_z$ (black is $-1$ , white is $+1$).}
\label{fig:vortices}
\end{figure}

In a thin film, shape anisotropy due to dipolar interactions forces magnetization $\mathbf m$ into the plane of the film, $m_z = 0$. A vortex possesses a small core where $m_z \neq 0$. Away from the core, the in-plane magnetization rotates through $+2\pi$ as the core is circumnavigated counterclockwise. In equilibrium, a vortex has an axially symmetric structure. In polar coordinates $(r, \alpha)$ with  $r = 0$ at the core, $m_z$ is a function of the radius $r$, whereas the azimuthal magnetization angle $\phi$ is a function of the azimuthal angle in configuration space $\alpha$: 
\begin{subequations}
\begin{eqnarray}
m_z(0) &=& p = \pm 1, 
\quad
m_z(\infty) = 0,
\label{eq:vortex-equilibrium-a}
\\
\phi(\alpha) &=& \alpha + \chi \pi/2 = \alpha \pm \pi/2,
\label{eq:vortex-equilibrium-b}
\end{eqnarray}
where $\chi = \pm 1$ is the ``chirality'' of the vortex and $p = \pm 1$ the ``polarity'' of its core. The locking of magnetization to the azimuthal direction in configuration space (\ref{eq:vortex-equilibrium-b}), is enforced by dipolar interactions \cite{Aharoni1990}. Two values of polarity and two of chirality give four distinct vortex configurations, Fig.~\ref{fig:vortices}. 
\label{eq:vortex-equilibrium}
\end{subequations}

\subsection{Canonical momenta}

Canonical momenta can be obtained from Eq.~(\ref{p-canonical-soliton}) along the lines of Sec.~\ref{sec:domain-wall-p}. Linear momentum is
\begin{equation}
p_\alpha = - \mathcal J 
	\int_\Omega \mathbf a(\mathbf m) \cdot \partial_\alpha \mathbf m \, d^2r.
\end{equation}
Here $\Omega$ is the two-dimensional area of the sample. Under a gauge transformation (\ref{eq:gauge-m}), linear momentum changes as follows: 
\begin{equation}
\mathbf p' 
	= \mathbf p - \int_\Omega \nabla_{\mathbf r} \chi \, d^2r
	= \mathbf p - \int_{\partial \Omega} \chi \hat{\mathbf n} \, dr.
\end{equation}
Here $\partial \Omega$ is the boundary of the sample and $\hat{\mathbf n}$ is its outward normal. Again, canonical momentum is gauge-dependent and unphysical.

In the standard gauges (\ref{eq:gauge-poles}), 
\begin{equation}
p_\alpha = - \mathcal J \int (\cos{\theta}\pm 1) \partial_\alpha \phi \, d^2r.
\label{eq:p-vortex-integral}
\end{equation}
The same expression results from the application of Noether's theorem \cite{Kosevich1990, Papanicolaou1991}. Evaluation of the integral in Eq.~(\ref{eq:p-vortex-integral}) requires care: the integrand decays slowly away from the core. To ensure its convergence, we restrict the integration to a disk of radius $R_d$ much larger than both the vortex core and the vortex displacement $\mathbf R$ and subtract the momentum of the vortex at the origin, $p_\alpha(\mathbf R) - p_\alpha(0)$ (see \ref{app:p-vortex}). Upon the subtraction, contributions from the interior of the disk cancel out, leaving behind an integral over a shell of width $R \ll R_d$ on the edge. The difference is linear in $\mathbf R$ and is independent of vortex polarity $p$ because the contributions from the core canceled out after the subtraction:
\begin{equation}
p_\alpha(\mathbf R) =  p_\alpha(0) \pm \pi \mathcal J \epsilon_{\alpha\beta} X_\beta,
\quad
\mathbf m_s = \pm \hat{\mathbf z}.
\end{equation}

To avoid the spurious contribution of the Dirac string to canonical momentum $p_\alpha$, we should attach the string where magnetization cannot reach it, i.e., opposite to the magnetization at the core, $\mathbf m_s = - p \hat{\mathbf z}$. Then 
\begin{equation}
p_\alpha(\mathbf R) =  p_\alpha(0) - p \pi \mathcal J \epsilon_{\alpha\beta} X_\beta.
\label{eq:p-vortex}
\end{equation}

We shall see shortly that these values are only half the correct answer, given below by Eq.~(\ref{eq:P-vortex}). As explained in Sec.~\ref{sec:mechanics-stuff}, this failure is unavoidable. That canonical momentum $\mathbf p$ is only half of conserved momentum $\mathbf P$ is consistent with Eqs.~(\ref{eq:p-q-relations}).

Like with the domain wall, it is possible to match one of the canonical momenta with its conserved counterpart---at the expense of the other. Place the Dirac string somewhere at the $\phi_s = 0$ meridian, $\mathbf m_s = (\sin{\theta_s}, 0, \cos{\theta_s})$. We then obtain (\ref{app:p-vortex})
\begin{eqnarray}
p_X(\mathbf R) &=& p_X(0) 
	+ \frac{2 \pi \, \mathrm{sgn}\cos{\theta_s}}{1+|\cos{\theta_s}|} \mathcal J Y, 
\nonumber\\
p_Y(\mathbf R) &=& p_Y(0)
	- \frac{2 \pi \, \cos{\theta_s}}{1+|\cos{\theta_s}|} \mathcal J X.
\label{eq:p-vortex-generic-string}
\end{eqnarray}
By placing the string just above or below the equator (depending on the polarity), $\cos{\theta_s} = \mp 0$ for $p = \pm 1$, we get $p_X = - 2 p \pi \mathcal J Y$ in accordance with Eq.~(\ref{eq:P-vortex}) below. However, $p_Y = 0$ is obviously wrong. 

\subsection{Conserved momenta}

To compute conserved momenta $P_X$ and $P_Y$ conjugate to rigid displacements $X$ and $Y$ of the vortex with the aid of Eq.~(\ref{eq:P-soliton}) for an undeformed vortex, we need the gyrotropic coefficients $F_{XY} = - F_{YX}$ only (all other modes are switched off). As $\mathbf R = (X,Y)$ represents rigid displacements of a magnetic texture, we replace $\partial_X = - \partial_x$ and $\partial_Y = - \partial_y$ to obtain the gyrotropic coefficient
\begin{equation}
F_{XY} = \mathcal J \int 
	\mathbf m \cdot 
		\left( 
			\partial_x \mathbf m \times \partial_y \mathbf m
		\right) \, d^2r
		= 4 \pi Q \mathcal J
\label{eq:F-XY}
\end{equation}
and the Poisson bracket 
\begin{equation}
\{P_X, P_Y\} = - F_{XY} = - 4 \pi Q \mathcal J.
\label{eq:Poisson-2-dimensions}
\end{equation}
Here $Q$ is the skyrmion charge with density 
\begin{equation}
q = \frac{1}{4\pi} \mathbf m \cdot (\partial_x \mathbf m \times \partial_y \mathbf m)
\label{eq:skyrmion-density}
\end{equation}
is its density. For a vortex with polarity $p$, 
\begin{equation}
Q = p/2 = \pm 1/2.
\label{eq:Q-vortex} 
\end{equation}

As with the domain wall, the gyrotropic coefficient for the two zero modes of a vortex is a topological invariant insensitive to the detailed structure of its core \cite{Huber1982, Guslienko2005}. We thus obtain 
\begin{equation}
P_\alpha(\mathbf R) =  P_\alpha(0) - 4\pi Q \mathcal J \epsilon_{\alpha\beta} X_\beta 
\label{eq:P-vortex}
\end{equation}
for an undeformed vortex. Here $\alpha$ and $\beta$ are Cartesian indices and $\epsilon_{\alpha\beta}$ is the Levi-Civita symbol in 2 dimensions, $\epsilon_{xy} = - \epsilon_{yx} = 1$. 

Let us also compute the integral of motion associated with rotational symmetry. In the presence of spin-orbit coupling and dipolar interactions, rotations must involve  
magnetization components $m_x$ and $m_y$ as well as spatial coordinates $x$ and $y$. To evaluate the conserved angular momentum of an undeformed vortex with a core centered at $\mathbf R = (X,Y)$, we need the gyrotropic coefficients $F_{\Phi X}$ and $F_{\Phi Y}$, or $F_{\Phi \alpha}$ for brevity. An infinitesimal rotation about the origin is equivalent to a global translation by $\delta X_\alpha = -\epsilon_{\alpha\beta} X_\beta \, \delta \Phi$ followed by a rotation about the vortex core. The latter operation does not affect the vortex because of its axial symmetry (Fig.~\ref{fig:vortices}). Thus a rotation reduces to a pure translation, giving a relation between the derivatives 
\begin{equation}
\frac{\partial \mathbf m}{\partial \Phi} 
	= \epsilon_{\alpha\beta} X_\alpha \frac{\partial \mathbf m}{\partial X_\beta}.
\end{equation}
The gyrotropic coefficients $F_{\Phi X}$ and $F_{\Phi Y}$ can then be expressed in terms of $F_{XY}$: 
\begin{equation}
F_{\Phi \alpha} 
	=  - X_\alpha F_{XY} = - 4\pi Q \mathcal J X_\alpha.
\end{equation}
We thus obtain the angular momentum of an undeformed vortex centered at $\mathbf R$: 
\begin{equation}
P_\Phi(\mathbf R) = P_\Phi(0) - \int F_{\Phi \alpha} \, dX_\alpha 
= P_\Phi(0) + 2 \pi Q \mathcal J R^2.
\end{equation}

\subsection{Conjecture of Papanicolaou and Tomaras}

Papanicolaou and Tomaras \cite{Papanicolaou1991} conjectured that linear momentum of a two-dimensional ferromagnet is given by the following expression: 
\begin{equation}
P_\alpha 
	= - 4\pi \mathcal J \epsilon_{\alpha\beta} \int x_\beta \, q
		\, d^2r, 
\label{eq:P-PT}
\end{equation}
Although they did not derive this result, they offered plausible arguments in favor of this conjecture. For example, they showed that $P_\alpha$ is the generator of translations for the field $\mathbf m(\mathbf r)$. Additionally, Sch\"utte and Garst \cite{Schutte2014} pointed out that translating a soliton by $\delta X_\alpha$ without deformation increments linear momentum (\ref{eq:P-PT}) by $\Delta P_\alpha = - 4\pi Q \mathcal J \epsilon_{\alpha \beta} \, \delta X_\beta$, in agreement with our Eq.~(\ref{eq:P-vortex}). 

Did Papanicolaou and Tomaras guess correctly the conserved momentum of a ferromagnet in two dimensions? The answer appears to be yes. The general expression  (\ref{eq:P-field}) gives the following result for linear momentum in two dimensions: 
\begin{equation}
P_\alpha(t) = P_\alpha(0) + \mathcal J \int_0^t dt \int 
	\mathbf m \cdot (\partial_\alpha \mathbf m \times \partial_t \mathbf m) \, d^2r. 
\label{eq:P-2-dimensions}
\end{equation}
The time derivatives $dP_\alpha/dt$ of momenta (\ref{eq:P-PT}) and (\ref{eq:P-2-dimensions}) differ by a boundary term 
\begin{equation}
 \epsilon_{\alpha\beta} \oint x_\beta \, 
	\mathbf m \cdot (\partial_\gamma \mathbf m \times \partial_t \mathbf m) \, 
	dx_\gamma.
\end{equation}
If magnetization at the boundary is confined to a single plane (e.g., constrained to lie in the plane of the film or to be normal to it) or if it is completely static then the boundary term vanishes, and the definition of Papanicolaou and Tomaras is equivalent to ours. Their conjecture is thus confirmed for reasonable boundary conditions. 

\section{Discussion}
\label{sec:discussion}

We have shown that the canonical recipe for computing conserved momenta through the application of Noether's theorem quite generally fails for a ferromagnet. The problem is brought into focus by a close analogy between the dynamics of a ferromagnetic soliton and of a charged particle in a magnetic field. In the latter case, it is well known that canonical momenta are gauge dependent and generally not conserved. The presence of a background gauge field makes it necessary to follow a physical symmetry with a gauge transformation. Conserved momenta are generators of these combined transformations. They are different from canonical momenta. The group of magnetic translations \cite{Zak1964} is one of the oldest examples of a gauged symmetry. It is frequently used in the context of the quantum Hall effect \cite{Xiao2010}. Gauged angular momentum goes even further back in time \cite{Fierz1944}. Wen's projective symmetry group \cite{Wen2002} is a relatively recent application of gauged symmetries.

We have exploited this analogy to properly define conserved momenta of ferromagnetic solitons. Eq.~(\ref{eq:P-soliton}) expresses them as a function of collective coordinates and Eq.~(\ref{eq:P-field}) as a functional of the magnetization field. Computing conserved momenta $P_a$ of a soliton in equilibrium (no deformations) is straightforward: it requires the knowledge of gyrotropic coefficients $F_{ab}$ \cite{Tretiakov2008, Clarke2008} of the soliton's zero modes, which are easy to calculate. 

Two well-known models of ferromagnetic solitons---a domain wall in a ferromagnetic chain and a vortex in a thin film---have been used to illustrate these general considerations. We have derived the conserved momenta and have shown that canonical momenta obtained through the na\"{\i}ve application of Noether's theorem generally fail to give the correct answer. 

Specializing to two dimensions, our results for conserved linear and angular momenta confirm an earlier conjecture by Papanicolaou and Tomaras \cite{Papanicolaou1991}. Thus our method resolves paradoxes accumulated over decades and provides a straightforward and very general method for computing conserved momenta of ferromagnetic solitons. 

It is worth noting that early works by Slonczewski \cite{Slonczewski1973} and by Thiele \cite{Thiele1976} contained the correct treatment of linear momentum of a domain wall. These researchers relied on Newton's second law for solitons and calculated linear momentum by integrating the impulse of an external force along the lines of Eqs.~(\ref{eq:Newton-2nd-law}) and (\ref{eq:magnetic-momentum}). They even anticipated ``the treacherous nature'' \cite{Thiele1976} of  canonical momentum (\ref{eq:p-vortex-integral})!

\begin{figure}
\includegraphics[width=0.95\columnwidth]{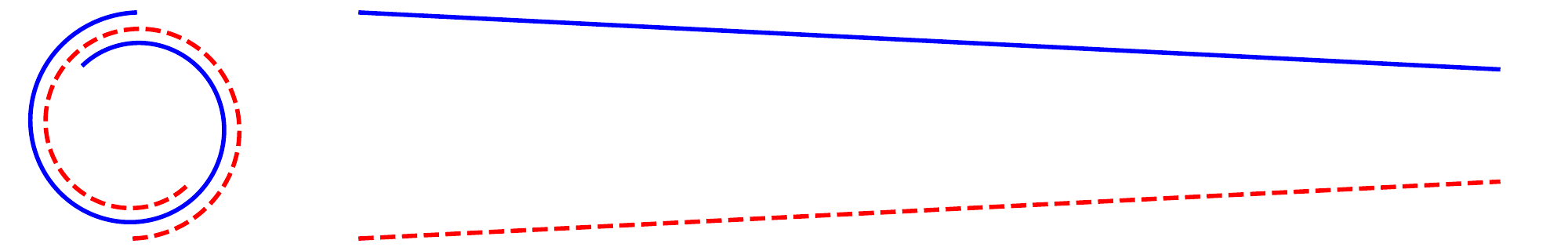}
\caption{Trajectories of an interacting pair of solitons, a vortex (blue solid line) and an antivortex (red dashed line). Left: equal skyrmion numbers, $Q_1 = Q_2$. Right: opposite skyrmion numbers, $Q_1 = - Q_2$.}
\label{fig:trajectories}
\end{figure}

The existence of conserved momenta can be useful in situations where solitons interact with one another. For example, if two vortices in a thin film (typically a vortex and an antivortex) attract each other, their net conserved linear momentum is 
\begin{equation}
\mathbf P = - 4 \pi \mathcal J \hat{\mathbf z} \times (Q_1 \mathbf R_1 + Q_2 \mathbf R_2).
\end{equation}
If the vortices have equal skyrmion numbers, $Q_1 = Q_2$, then their average position remains in place, $(\mathbf R_1 + \mathbf R_2)/2 = \mathrm{const}$, which means that they orbit a common center. If the skyrmion numbers are equal and opposite, $Q_1 = - Q_2$, then their relative position remains fixed, $\mathbf R_1 - \mathbf R_2 = \mathrm{const}$, as the two vortices move in the direction orthogonal to the line connecting them. The two situations are illustrated in Fig.~\ref{fig:trajectories}. The gradual convergence of the two solitons in the figure reflects the influence of weak viscous friction. 

Similarly, two consecutive domain walls in a ferromagnetic wire (Sec.~\ref{sec:domain-wall}) have the total linear momentum $P_Z = 2 \mathcal J(\Phi_1 - \Phi_2)$. In the absence of external forces, the two interacting domain walls will precess at the same frequency but keep their relative orientation fixed, $\Phi_1 - \Phi_2 = \mathrm{const}$.  

\section*{Acknowledgments} 

I thank Gerrit Bauer, Markus Garst, Boris Ivanov, Se Kwon Kim, Masaki Oshikawa, Denis Sheka, Yasuhiro Tada, Ari Turner, and Haruki Watanabe for stimulating discussions. I gratefully acknowledge hospitality of the Institute for Solid State Physics of the University of Tokyo and of the Kavli Institute for Theoretical Physics. This work was supported by the Japan Society for the Promotion of Science, by the U.S. Department of Energy, Office of Basic Energy Sciences, Division of Materials Sciences and Engineering under Award DE-FG02-08ER46544, and by the National Science Foundation under Grant No. NSF PHY11-25915.

\appendix

\section{Magnetic field $F_{ij}$ from gauge potential $A_i$}
\label{appendix:F-from-A}

The magnetic field $F_{ij}$ felt by a soliton (\ref{eq:F-gyrotropic}) can be derived in a number of ways. Tretiakov \emph{et al.} \cite{Tretiakov2008} obtain it by translating the Landau-Lifshitz equation of motion for the magnetization field $\mathbf m(\mathbf r,t)$ into the language of collective coordinates $\mathbf q(t)$. Here we obtain it directly from the definition of the gauge potential (\ref{eq:L-soliton}) and magnetic field (\ref{eq:F-def}): 
\begin{eqnarray}
F_{ij} &=& \frac{\partial A_j}{\partial q_i} - \frac{\partial A_i}{\partial q_j}
= \int  
	\left[ 
		\frac{\partial a_\alpha}{\partial q_i} \frac{\partial m_\alpha}{\partial q_j}
		- (i \leftrightarrow j) 
	\right]
\, dV
\nonumber\\
&=& 
\int  
	\left[
		\frac{\partial a_\alpha}{\partial m_\beta} 
		\frac{\partial m_\beta}{\partial q_i}
		\frac{\partial m_\alpha}{\partial q_j} 
		- (i \leftrightarrow j)
	\right] 
\, dV
= \int  
	\left(
		\frac{\partial a_\alpha}{\partial m_\beta} 
		- \frac{\partial a_\beta}{\partial m_\alpha}
	\right)
	\frac{\partial m_\beta}{\partial q_i} 
	\frac{\partial m_\alpha}{\partial q_j}
\, dV
\nonumber\\
&=& 
	- \int 
		\mathcal J \epsilon_{\alpha\beta\gamma} m_\gamma
		\frac{\partial m_\beta}{\partial q_i} 
		\frac{\partial m_\alpha}{\partial q_j}
\, dV
= \mathcal J \int \mathbf m \cdot 
		\left( 
			\frac{\partial \mathbf m}{\partial q_i}
			\times 
			\frac{\partial \mathbf m}{\partial q_j}
		\right) \, 
		dV.
\end{eqnarray}
In this derivation we relied on Eq.~(\ref{eq:a-curl}), which states that the spin vector potential $\mathbf a(\mathbf m)$ describes a magnetic monopole, $\nabla_{\mathbf m} \times \mathbf a = - \mathcal J \mathbf m$, or
\begin{equation}
\frac{\partial a_\alpha}{\partial m_\beta} - \frac{\partial a_\beta}{\partial m_\alpha}
= - \mathcal J \epsilon_{\alpha\beta\gamma} m_\gamma.
\end{equation}

\section{Poisson brackets for collective coordinates and conserved momenta of a soliton}
\label{appendix:Poisson}

We first establish that the tensor of Poisson brackets for collective coordinates $\{q_i, q_j\} = Q_{ij}$ is the inverse of the gyrotropic tensor $F_{ij}$ (\ref{eq:F-gyrotropic}): $F Q = 1$, i.e., $F_{ij} Q_{jk} = \delta_{ik}$. (Summation over doubly repeated indices is implied.)
\begin{eqnarray}
F_{ij} Q_{jk} &=& \iint dV \, dV' \, 
	\left(
		\frac{\partial \phi}{\partial q_i} \frac{\partial \cos{\theta}}{\partial q_j}
		-  \frac{\partial \cos{\theta}}{\partial q_i} \frac{\partial \phi}{\partial q_j}
	\right)
	\left( 
		\frac{\delta q_j}{\delta \cos{\theta'}} \frac{\delta q_k}{\delta \phi'}
		- \frac{\delta q_j}{\delta \phi'} \frac{\delta q_k}{\delta \cos{\theta'}} 
	\right)
\nonumber\\
	&=& \int dV \, 
	\left(
		\frac{\partial \phi}{\partial q_i} 
			\frac{\delta q_k}{\delta \phi}
		+ \frac{\partial \cos{\theta}}{\partial q_i}
			\frac{\delta q_k}{\delta \cos{\theta}}
	\right)
	= \frac{\partial q_k}{\partial q_i} = \delta_{ik}.
\end{eqnarray}
Here we used the differentiation chain rule and functional derivatives 
\begin{equation}
\frac{\delta \cos{\theta}}{\delta \cos{\theta'}} 
	= \frac{\delta \phi}{\delta \phi'} 
	= \delta(\mathbf r - \mathbf r'),
\quad
\frac{\delta \cos{\theta}}{\delta \phi'} 
	= \frac{\delta \phi}{\delta \cos{\theta'}} = 0.
\end{equation}
We thus obtain the Poisson bracket for two collective coordinates in Eq.~(\ref{eq:Poisson-soliton}), 
\begin{equation}
\{q_i, q_j\} = (F^{-1})_{ij}.
\end{equation}

The remaining two results in Eq.~(\ref{eq:Poisson-soliton}) can be obtained by using the  identity 
\begin{equation}
\{f,g\} 
	= \frac{\partial f}{\partial q_i} 
		\frac{\partial g}{\partial q_j} 
		\{q_i, q_j\}
\label{eq:Poisson-soliton-q-q}
\end{equation}
valid for arbitrary functions $f(\mathbf q)$ and $g(\mathbf q)$. The mixed Poisson bracket in Eq.~(\ref{eq:Poisson-soliton}) is
\begin{equation}
\{P_a, q_i\} 
	= \frac{\partial P_a}{\partial q_j} 
		\frac{\partial q_i}{\partial q_k} 
		\{q_j,q_k\}
	= -F_{aj} \delta_{ik} (F^{-1})_{jk} = - \delta_{ai},
\end{equation}
where the identity $\partial P_a / \partial q_j = - F_{aj}$ follows from the definition of conserved momentum (\ref{eq:P-soliton}). 

The Poisson bracket for two conserved momenta in Eq.~(\ref{eq:Poisson-soliton}) is obtained along the same lines: 
\begin{equation}
\{P_a, P_b\} 
	= \frac{\partial P_a}{\partial q_i} 
		\frac{\partial P_b}{\partial q_j}
		\{q_i,q_j\}
	= F_{ai} F_{bj} (F^{-1})_{ij} = - F_{ab}. 
\end{equation}

\section{Canonical momentum of a vortex}
\label{app:p-vortex}

We evaluate the canonical momentum $\mathbf p$ of a vortex, Eq.~(\ref{eq:p-vortex-integral}). The integrand decays slowly away from the core: $\cos{\theta} \pm 1 \to \pm 1$ and $\partial_\alpha \phi \sim - \epsilon_{\alpha\beta} x_\beta/r^2$. To ensure convergence, we restrict the integration to a finite disk of radius $R_d$ greatly exceeding both the core size and displacement $\mathbf R$: 
\begin{equation}
p_\alpha(\mathbf R) 
	= - \mathcal J \int_{r<R_d} 
		(\cos{\theta}\pm 1) \partial_\alpha \phi \, d^2r.
\end{equation} 

Next we subtract the momentum of a vortex with the core at the center of the disk, whose configuration is $\mathbf m_0(\mathbf r)$:
\begin{eqnarray}
p_\alpha(\mathbf R) - p_\alpha(0) 
	&=& - \mathcal J \int_{|\mathbf r-\mathbf R|<R_d} 
		(\cos{\theta_0}\pm 1) \partial_\alpha \phi_0 \, d^2r
\nonumber\\
	&& + \mathcal J \int_{r<R_d} 
		(\cos{\theta_0}\pm 1) \partial_\alpha \phi_0 \, d^2r.
\end{eqnarray} 
The contributions from the area where the disks $r<R_d$ and $|\mathbf r-\mathbf R|<R_d$ overlap cancel out and we are left with an integral over two crescents of thickness $R$ bounded by the edges of the disks. For a disk radius $R_d$ greatly exceeding the core size, we may set $\cos\theta_0 \to 0$ in the crescents. 

The integration over $\mathbf r$ is conveniently done in polar coordinates, radius $r$ and direction $\hat{\mathbf n} = \mathbf r/r = (\cos{\psi}, \sin{\psi})$. For $R_d \gg R$, $\partial_\alpha \phi_0 = \epsilon_{\alpha\beta} n_\beta/r$ is approximately constant across the thickness of the crescent, so that for a given $\hat{\mathbf n}$, the integration over $r$ yields a factor of $\hat{\mathbf n} \cdot \mathbf R = n_\gamma X_\gamma$. Integration over directions yields Eq.~(\ref{eq:p-vortex}):
\begin{equation}
p_\alpha(\mathbf R) - p_\alpha(0) 
	= \pm \mathcal J \int_0^{2\pi}  
		\epsilon_{\alpha\beta} n_\beta n_\gamma X_\gamma 
		\, d \psi
	= \pm \pi \mathcal J \epsilon_{\alpha\beta} X_\beta, 
\quad
\mathbf m_s = \pm \hat{\mathbf z}.
\end{equation} 

Lastly, we derive the canonical linear momentum when the Dirac string is attached at a generic location along the $\phi_s = 0$ meridian, $\mathbf m_s = (\sin{\theta_s}, 0, \cos{\theta_s})$, with the gauge (\ref{eq:gauge-generic}). We use the same approach as above, introducing a long-distance cutoff $R_d$ and evaluating the momentum difference $\mathbf p(\mathbf R) - \mathbf p(0)$. Doing so again reduces the integration area to two narrow crescents of radius $R_d$ and width $R$. Since the integration area is far away from the core, magnetization lies in the $xy$ plane and 
\begin{equation}
\mathbf m_s \cdot (\mathbf m_0 \times \partial_\alpha \mathbf m_0)   
	= \mathbf m_s \cdot \hat{\mathbf z} \, \partial_\alpha \phi_0
	= - \cos{\theta_s} \epsilon_{\alpha\beta} n_\beta/R_d.
\end{equation}
We thus obtain Eq.~(\ref{eq:p-vortex-generic-string}):
\begin{eqnarray}
p_X(\mathbf R) - p_X(0) 
	&=& +\mathcal J \int_0^{2\pi} 
		\frac{\cos{\theta_s} \sin{\psi}(X \cos{\psi + Y \sin{\psi}})}
			{1 + \sin{\theta_s} \sin{\psi}}
		\, d\psi
\nonumber\\
	&=& \frac{2\pi \, \mathrm{sgn}\cos{\theta_s}}{1+|\cos{\theta_s}|} \mathcal J Y,
\nonumber\\
p_Y(\mathbf R) - p_Y(0) 
	&=& -\mathcal J \int_0^{2\pi} 
		\frac{\cos{\theta_s} \cos{\psi}(X \cos{\psi + Y \sin{\psi}})}
			{1 + \sin{\theta_s} \sin{\psi}}
		\, d\psi
\nonumber\\
	&=& - \frac{2\pi \cos{\theta_s}}{1+|\cos{\theta_s}|} \mathcal J X.
\end{eqnarray}


\bibliographystyle{elsarticle-num}
\bibliography{momentum}

\end{document}